\documentclass[symmetry,article,accept,moreauthors,pdftex,accept]{mdpi} 
\usepackage{color} 
\usepackage{graphics}
\usepackage{graphicx}
\usepackage{amssymb}
\usepackage{color}
\usepackage{amsfonts}
\usepackage{caption}
\usepackage{hyperref}
\usepackage{hyperref}
\usepackage[hypcap=true]{caption}
\usepackage{caption}
\usepackage[labelformat=simple]{subcaption}

\firstpage{1} 
\pubvolume{xx}
\issuenum{1}
\articlenumber{5}
\pubyear{2019}
\copyrightyear{2019}
\history{Received: 22 July, 2019;  Accepted: 18 September 2019; Published: date}
\updates{yes} % If there is an update available, un-comment this line

\Title{Behavior of Floquet Topological Quantum States in Optically Driven Semiconductors}

\Author{Andreas Lubatsch $^{1\dagger}$ and Regine Frank $^{2,3\dagger}$*}

\AuthorNames{Andreas Lubatsch  and Regine Frank}

\address{%
$^{1}$ \quad Georg-Simon-Ohm University of Applied Sciences, Ke{\ss}lerplatz
  12, 90489 N\"urnberg, Germany\\
$^{2}$ \quad Bell Laboratories, 600 Mountain Avenue, Murray Hill, NJ 07974-0636, USA\\
$^{3}$ \quad Serin Physics Laboratory, Department of Physics and Astronomy,
Rutgers University, 136 Frelinghuysen Road, Piscataway, NJ 08854-8019, USA; regine.frank@rutgers.edu}

\corres{Correspondence: regine.frank@googlemail.com, regine.frank@rutgers.edu}

\firstnote{These authors contributed equally to this work.}
\abstract{Spatially uniform optical excitations can induce Floquet topological
  band structures within insulators which can develop similar or equal
  characteristics as are known from three-dimensional topological insulators. We derive in this article theoretically
  the development of Floquet topological quantum states for
  electromagnetically driven semiconductor bulk matter and we present results
  for the lifetime of these states and their occupation in the
  non-equilibrium. The direct
  physical impact of the mathematical precision of the Floquet-Keldysh
  theory is evident when we solve the driven system of a generalized Hubbard model with
  our framework of dynamical mean field theory (DMFT) in the non-equilibrium
  for a case of ZnO. The physical
  consequences of the topological non-equilibrium effects in our results for
  correlated systems are explained with their impact on optoelectronic applications.}


\begin{document}
%%%%%%%%%%%%%%%%%%%%%%%%%%%%%%%%%%%%%%%%%%

%%%%%%%%%%%%%%%%%%%%%%%%%%%%%%%%%%%%%%%%%%
\setcounter{section}{-1} %% Remove this when starting to work on the template.

\section{Introduction}
\label{INTRO}

Topological phases of matter \cite{Kosterlitz,KaneMele,HasanKane} have captured our fascination
over the past decades, revealing properties in the sense of robust edge modes
and exotic non-Abelian excitations \cite{Fu,Moore}. Potential applications of
periodically driven quantum systems \cite{Dalibard} are
conceivable in the subjects of semiconductor spintronics \cite{Zuti} up to topological quantum
computation \cite{Nayak} as well as topological lasers
\cite{Rechtsman,Demetrios} in optics and
random lasers \cite{LubatschAPPLSCI}. Already topological insulators in solid-state devices such as
HgTe/CdTe quantum wells \cite{Bernevig,Knig}, as well as topological Dirac insulators
such as Bi$_2$Te$_3$ and Bi$_2$Sn$_3$ \cite{Hsieh,Xia,Zhang} were
groundbreaking discoveries
in the search for the unique properties of topological phases and their technological applications.\\ 
In non-equilibrium systems it has been shown that time-periodic
perturbations can induce topological properties in conventional insulators
\cite{Lindner1,Podolsky,Lindner2,Wang} which are trivial in equilibrium otherwise. Floquet topological
insulators include a very broad range of physical solid state and atomic realizations,
driven at resonance or off-resonance. These systems
can display metallic conduction which is enabled by quasi-stationary states
at the edges \cite{Kitagawa,Gu,Lindner1}. Their band structure may have the
form of a  Dirac cone in three dimensional
systems \cite{Gedik,Lindner3}, and Floquet Majorana fermions
\cite{Zoller} have been conceptionally developed. Graphene and Floquet
fractional Chern insulators have been recently investigated \cite{Rigol,Neupert,Bergholtz}.\\

\begin{figure}
\centering\resizebox{0.8\textwidth}{!}{%
  \includegraphics{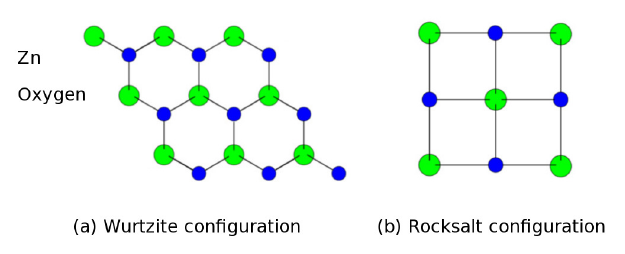}
}
%\vspace{0.5cm}       % Give the correct figure height in cm
\caption{ZnO structure (ab-plane). {\bf (a)} Non-centrosymmetric, hexagonal, wurtzite
    configuration {\bf (b)}
    Centrosymmetric, cubic, rocksalt configuration (Rochelle salt)
    \cite{JMATCHEMA,Rocksalt2,Rocksalt1}. The rocksalt configuration is
    distinguished by a tunable gap from $1.8\,eV$ up to $6.1\,eV$, a gap value
    of $2.45\,eV$ is typical for the  monocrystal rocksalt configuration without
    oxygen vacancies \cite{ZnOMono1,ZnOMono2}. As such the rocksalt
    configuration could be suited for higher harmonics generation
    under non-equilibrium topological excitation \cite{Faisal0,LubatschEPJ2019}.}
\label{ZSTRUCT}       % Give a unique label
\end{figure}

In this article we show that Floquet topological quantum states can evolve in
correlated electronic systems of driven semi-conductors in the
non-equilibrium. We investigate $ZnO$ bulk matter in the centrosymmetric, cubic rocksalt
configuration, see Fig.(\ref{ZSTRUCT}). The non-equilibrium is
in this sense defined by the intense external electromagnetic driving field which
induces topologically dressed electronic states and the evolution of dynamical
gaps. These procedures are expected to be observable in pump-probe experiments
on time scales below the thermalization time. We
show that the expansion into Floquet modes
\cite{Floquet} is leading to results of direct physical impact in the sense of
modeling the coupling of a classical electromagnetic external
driving field to the correlated quantum many body system. Our results derived by
Dynamical Mean Field Theory (DMFT) in the non-equilibrium provide novel
insights in topologically induced phase transitions of driven otherwise
conventional 3-dimensional semiconductor bulk matter and insulators.

\section{Quantum Many Body Theory for Correlated Electrons in the Non-Equilibrium}
\label{sec:HUBBARDMODEL}

\begin{figure}
\centering\resizebox{0.6\textwidth}{!}{%
  \includegraphics{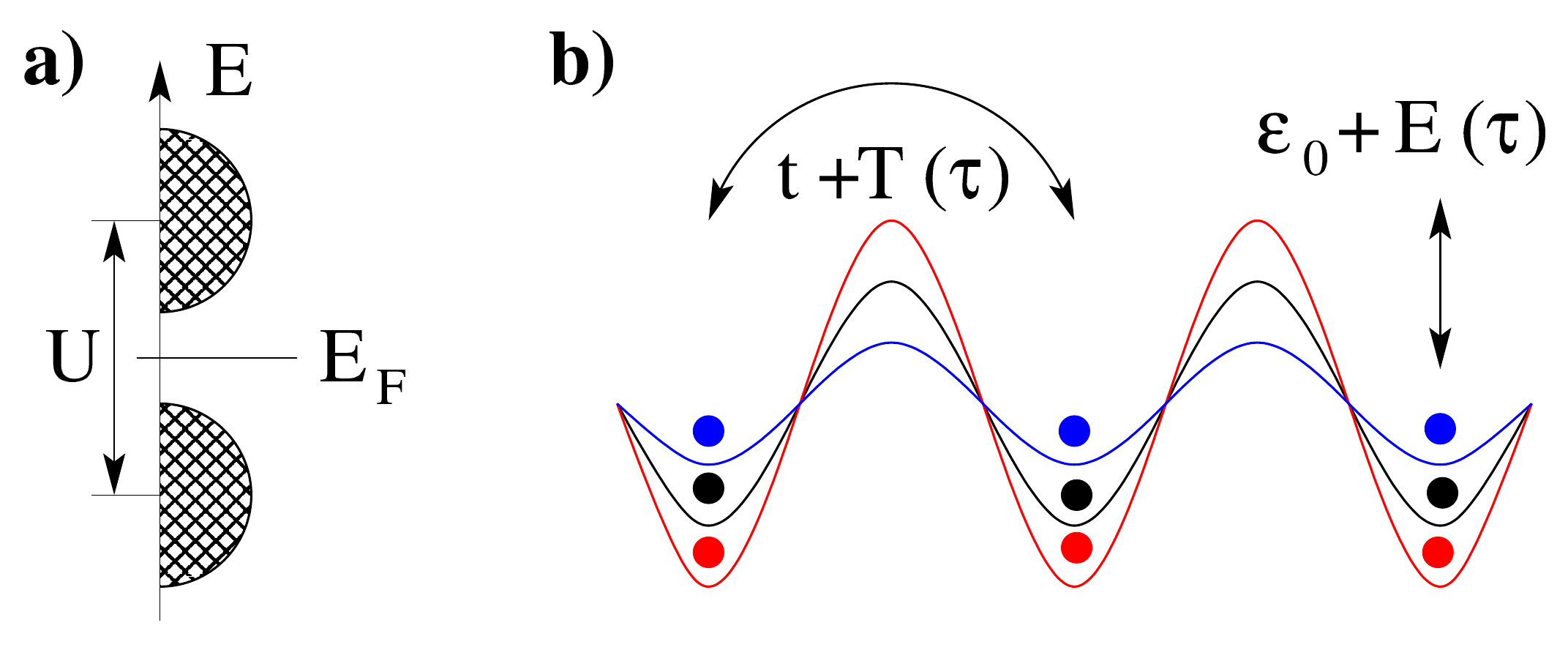}
}
%\vspace{0.5cm}       % Give the correct figure height in cm
\caption{Insulator to metal transition caused by photo-excitation. 
(a) Schematic split of energy bands due to the local Coulomb interaction
$U$. The gap is determined symmetrically to the Fermi edge $E_F$. 
(b)  The periodic in time driving yields an
additional hopping contribution $T(\tau)$ of electrons on the lattice (black) and the renormalization of the local potential, $E(\tau)$, as a quasi-energy. 
Colors of the lattice potential represent the external driving in time.}
\label{Mott}       % Give a unique label
\end{figure}

\begin{figure}
\centering\resizebox{0.7\textwidth}{!}{\includegraphics{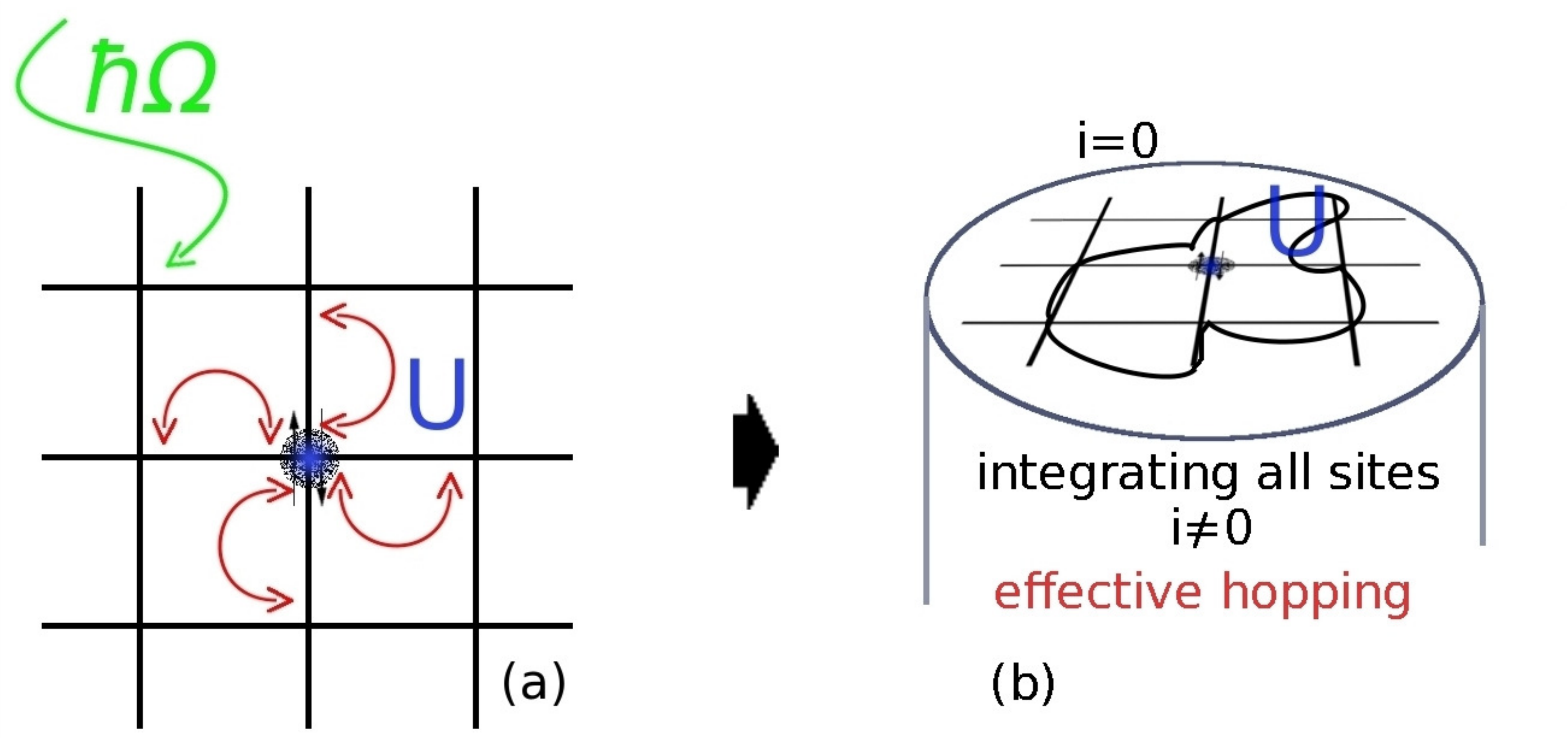}}
\vspace{0.5cm}       % Give the correct figure height in cm
\caption{Schematic representation of non-equilibrium dynamical mean field theory. 
(a) The semiconductor behaves in the here considered regime as an insulator:
Optical excitations by an external electromagnetic field  with the energy
$\hbar \Omega$ yield additional hopping processes. These processes are mapped
onto the interaction with the single site on the background of the surrounding
lattice bath in addition to the regular kinetic processes and in addition to on-site Coulomb repulsion. 
(b) The idea of DMFT: The integration over all lattice sites leads to
an effective theory including non-equilibrium excitations. The bath consists of
all single sites and the approach is thus self-consistent. 
The driven electronic system may in principal couple to a surface-resonance or
an edge state. The coupling to these states can be enhanced by the external
excitation.}
\label{DMFT}       % Give a unique label
\end{figure}

We consider in this work the wide gap semiconductor bulk to be driven by a
strong periodic-in-time external field in the optical range which yields
higher-order photon absorption processes. The electronic dynamics of the
photo-excitation processes, see Fig. \ref{Mott}, is theoretically modelled by
a generalized, driven, Hubbard Hamiltonian, see Equation (\ref{Hamilton_we}). The system
is solved with a Keldysh formalism including the electron-photon interaction
in the sense of the coupling of the classical electromagnetic field to the electronic dipole and thus to the
  electronic hopping. This yields an additional kinetic contribution. We
  solve the system by the implementation of a
  dynamical mean field theory (DMFT), see Fig. \ref{DMFT}, with a generalized iterative perturbation theory solver (IPT), see
  Fig. \ref{IPT-Sigma}. The full interacting Hamiltonian,
  Equation (\ref{Hamilton_we}), is introduced as follows

\begin{eqnarray}
\!\!\!\!\!\!\!\!H\!&=&\!\! \sum_{i, \sigma} \! \varepsilon_i  c^{\dagger}_{i,\sigma}c^{{\color{white}\dagger}}_{i,\sigma} 
+   \frac{U}{2} \sum_{i, \sigma} c^{\dagger}_{i,\sigma}c_{i,\sigma}c^{\dagger}_{i,-\sigma}c_{i,-\sigma}\label{Hamilton_we}
\\&& - t\!\! \sum_{\langle ij \rangle, \sigma}\!\!
c^{\dagger}_{i,\sigma}c^{{\color{white}\dagger}}_{j,\sigma}\nonumber
\\&& + i\vec{d}\cdot\vec{E}_0 \cos(\Omega_L \tau)\sum_{<ij>,\sigma} 
 \left(
           c^{\dagger}_{i,\sigma}c^{{\color{white}\dagger}}_{j,\sigma} 
 	  -
           c^{\dagger}_{j,\sigma}c^{{\color{white}\dagger}}_{i,\sigma} \right).\nonumber
\end{eqnarray}

In our notation (\ref{Hamilton_we})  $c^\dagger,(c)$ are the creator (annihilator) of an electron. The subscripts $i,j$ indicate the site, $\langle i,j \rangle$ implies the sum over
nearest neighboring sites. \\ The term $\frac{U}{2} \sum_{i,
  \sigma}c^{\dagger}_{i,\sigma}c_{i,\sigma}c^{\dagger}_{i,-\sigma}c_{i,-\sigma}$
results from the repulsive onsite Coulomb interaction $U$ between electrons with
opposite spins. The third
term $-t \sum_{\langle ij \rangle, \sigma}\!\!
c^{\dagger}_{i,\sigma}c^{{\color{white}\dagger}}_{j,\sigma}$ describes the
standard hopping processes of electrons with the amplitude $t$ between nearest neighboring sites. Those
contributions form the standard Hubbard model which is generalized for our
purposes in what follows. The first term $\sum_{i, \sigma} \! \varepsilon_i
c^{\dagger}_{i,\sigma}c^{{\color{white}\dagger}}_{i,\sigma}$   generalizes the
Hubbard model with respect to the onsite energy, see Fig.(\ref{Mott}). The electronic
on-site energy is noted as $\varepsilon_i$. 
The external time-dependent electromagnetic driving is described in terms
of the field $\vec{E}_0$ with laser frequency $\Omega_L $, $\tau$ which
couples to the
electronic dipole $\hat{d}$ with strength $|\bf{d}|$ .
The expression $ i\vec{d}\cdot\vec{E}_0 \cos(\Omega_L \tau)\sum_{<ij>,\sigma} 
 \left(
           c^{\dagger}_{i,\sigma}c^{{\color{white}\dagger}}_{j,\sigma} 
 	  -
           c^{\dagger}_{j,\sigma}c^{{\color{white}\dagger}}_{i,\sigma}
         \right)$ describes the renormalization of the standard electronic
         hopping processes, as one possible contribution $T(\tau)$ in Fig. (\ref{Mott}),
         due to external influences.

\subsection{FLOQUET STATES - COUPLING Of A CLASSICAL DRIVING FIELD TO A QUANTUM DYNAMICAL SYSTEM}
\label{sec:FLOTHEORY}

By introducing the explicit time
dependency of the external field we solve the generalized Hubbard Hamiltonian,
Equation (\ref{Hamilton_we}). It yields Green's
functions which depend on two separate time arguments which are are Fourier
transform to frequency coordinates. These frequencies are chosen as the relative and the center-of-mass frequency \cite{PRB,FrankANN} and we
introduce an expansion into Floquet modes

\begin{eqnarray}
\label{Floquet-Fourier}
G_{mn}^{\alpha\beta} (\omega) &=&
\left\lmoustache \!\!{\rm d}{\tau_1^\alpha}\!{\rm d}{\tau_2^\beta}\right.
e^{-i\Omega_L(m{\tau_1^\alpha}-n{\tau_2^\beta})}
e^{i\omega({\tau_1^\alpha}-{\tau_2^\beta})}
G (\tau_1^\alpha,\tau_2^\beta)\nonumber\\
&\equiv&
G^{\alpha\beta} (\omega-m\Omega_L, \omega - n\Omega_L).
\end{eqnarray}

\begin{figure}
\centering\resizebox{0.7\textwidth}{!}{%
  \includegraphics{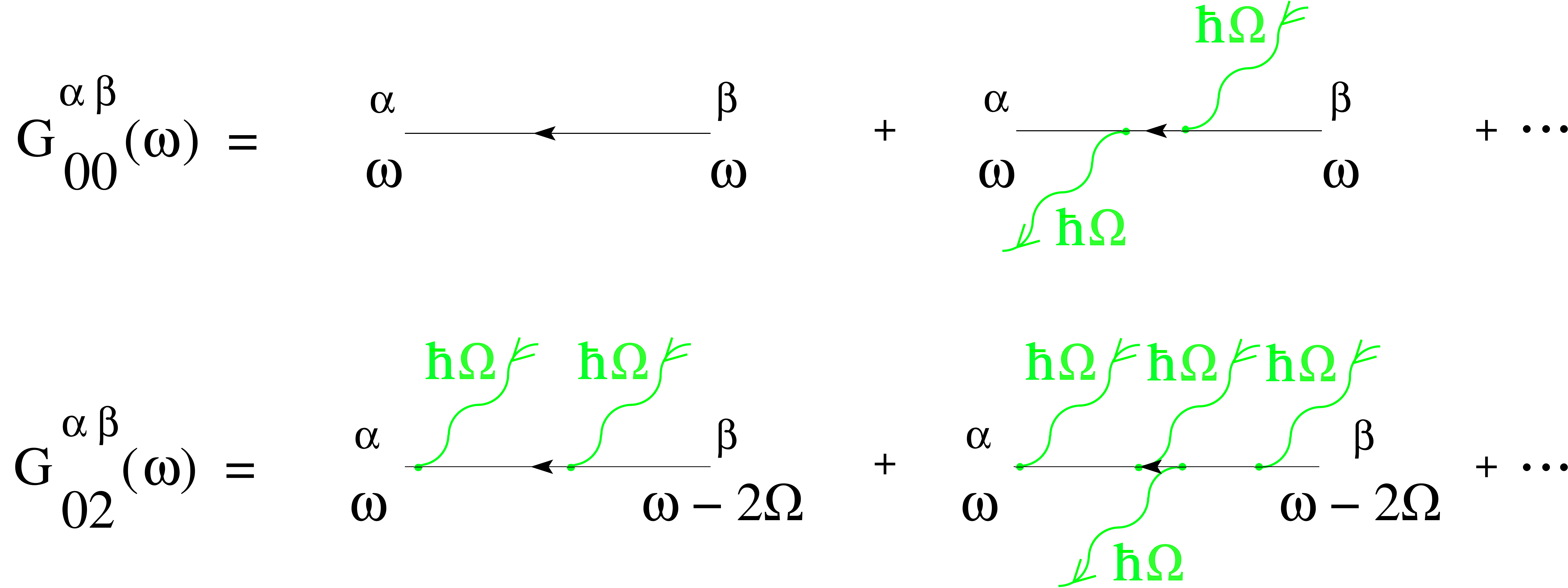}}
\vspace{0.5cm}       % Give the correct figure height in cm
\caption{Schematic representation of  the Floquet Green’s function and the Floquet matrix in terms of absorption and emission of
  external energy quanta $\hbar\Omega$. $G^{\alpha\, \beta}_{00} (\omega)$ represents the sum of all
  balanced contributions, $G^{\alpha\, \beta}_{02} (\omega)$ describes the net absorption of
  two photons.  $\alpha, \beta$ are the Keldysh indices.}
\label{FloquetGreen}       % Give a unique label
\end{figure}

In general Floquet \cite{Floquet} states are analogues to Bloch
states. Whereas Bloch states are due to the periodicity of the potential in
space, the spatial topology, the Floquet states represent the temporal
topology in the sense of the temporal periodicity \cite{Faisal0,Grifoni,Brandes,Eckardt,Uhrig,Sentef,Fan,PRB,FrankANN,ANN}. The Floquet expansion is introduced
in Fig. \ref{FloquetGreen} as a direct graphic representation of what is
described in Equation (\ref{Floquet-Fourier}). The Floquet modes are labelled by
the indices $(m,n)$, whereas $(\alpha,
\beta)$ refer to the branch of the Keldysh contour ($\pm$) and the
respective time argument. The physical consequence of the Floquet expansion
however is noteworthy, since it can be understood as the quantized absorption
and emission of energy $\hbar \Omega_L$ by the driven quantum many body system out of and into the classical external driving field.

In the case of uncorrelated electrons, $U\,=\,0$, the Hamiltonian can be
solved analytically and the retarded component of the Green's
function  $G_{mn}(k,\omega) $ reads 

\begin{eqnarray}
G_{mn}^{R}(k,\omega) \label{Green}
=
\sum_{\rho}
\frac
{
J_{\rho-m}\left(A_0\tilde{\epsilon}_k \right)
J_{\rho-n}\left(A_0\tilde{\epsilon}_k \right)
}
{
\omega -\rho\Omega_L - \epsilon_k + i 0^+
}.
\end{eqnarray}

Here $\tilde{\epsilon}_k$ is the dispersion relation  induced
by the external driving field. $\tilde{\epsilon}_k$ is to be distinguished from the
lattice dispersion $\epsilon$. $J_n$ are the cylindrical Bessel functions of
integer order, $A_0 = \vec{d}\cdot\vec{E}_0 $, 
$\Omega_L$ is the external laser frequency. The retarded Green's function for
the optically excited band electron is eventually given by

\begin{eqnarray}
\label{EqGsum}
G^{R}_{\rm Lb}(k,\omega)  
=
\sum_{m,n}
G_{mn}^{R}(k,\omega).
\end{eqnarray}

\subsection{DYNAMICAL MEAN FIELD THEORY IN THE NON-EQUILIBRIUM}
\label{sec:NUMSOL}

\begin{figure}
\hspace*{0.0cm}\centering\resizebox{0.6\textwidth}{!}{%
  \includegraphics{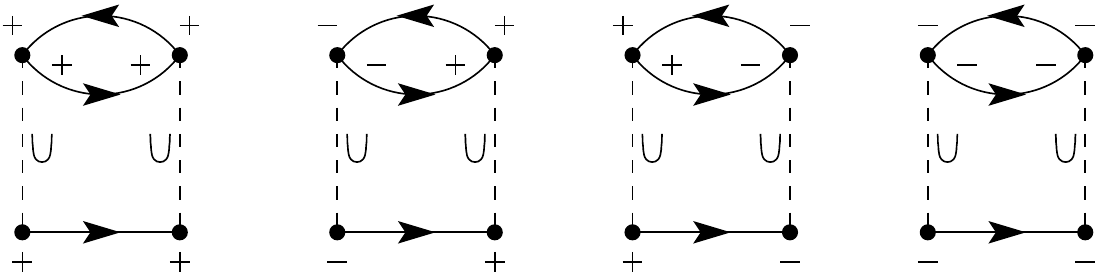}
}
\vspace{0.5cm}       % Give the correct figure height in cm
\caption{Local self-energy $\Sigma^{\alpha \beta}$ of the iterated
  perturbation theory (IPT). The IPT as a second order
  diagrammatic solver with respect to the electron electron interaction $U$
  is here generalized to non-equilibrium, $\pm$ indicates the branch of the Keldysh contour. 
The solid lines represent the bath in the sense of the Weiss-field
$\mathcal{G}^{\alpha\beta}$, see ref. \cite{NJP}.}
\label{IPT-Sigma}       % Give a unique label
\end{figure}

The generalized Hubbard model for the correlated system, $U\,\neq\,0$,
in the non-equilibrium, Equation (\ref{Hamilton_we}), is numerically solved by a single-site Dynamical Mean Field Theory (DMFT)
\cite{ANN,Kotliar,Metzner,Monien,Hettler,Freericks,OKA,Reichmann,Zgid,Werner,APLB,NJP,Max,Walter1,Walter}.  The expansion
into Floquet modes with the proper Keldysh description
models the external time dependent classical driving field, see \ref{sec:FLOTHEORY}, and couples it to the
quantum many body system. We numerically solve the Floquet-Keldysh DMFT 
\cite{ANN,NJP} with a second order iterative perturbation theory (IPT), where
the the local self-energy $\Sigma^{\alpha\beta}$ is derived by four bubble
diagrams, see Fig. \ref{IPT-Sigma}. The Green's function for the interaction of the laser with the 
band electron $G^{R}_{\rm Lb}(k,\omega)$, Equation (\ref{EqGsum}), is 
characterized by the wave vector $k$, where $k$ describes the periodicity of the
lattice. It  depends on the electronic frequency
$\omega$ and the external driving frequency $\Omega_L$, see Equation \ref{Floquet-Fourier}, captured in the Floquet indices $(m,n)$. The DMFT self-consistency relation, assumes the form of a matrix equation of non-equilibrium Green's
functions which is of dimension $2 \times 2$ in regular Keldysh space and of
dimension $n \times n$ in Floquet space.  The numerical algorithm is
efficient and stable also for all values of the Coulomb interaction $U$.

In previous work \cite{ANN,NJP,APLB} we considered an additional kinetic
energy contribution due to a lattice vibration. Here we take into account  a
coupling of the microscopic electronic dipole moment to an external
electromagnetic field \cite{PRB,FrankANN} for a correlated system. We
introduce the quantum-mechanical expression for the electronic dipole operator  $\hat d$, see
the last term r.h.s. Equation (\ref{Hamilton_we}), and this coupling reads as $ i\vec{d}\cdot\vec{E}_0 \cos(\Omega_L \tau)\sum_{<ij>,\sigma} 
 \left(
           c^{\dagger}_{i,\sigma}c^{{\color{white}\dagger}}_{j,\sigma} 
 	  -
           c^{\dagger}_{j,\sigma}c^{{\color{white}\dagger}}_{i,\sigma}
         \right)$. This kinetic contribution is conceptionally different from the generic kinetic
hopping of the third term of Equation (\ref{Hamilton_we}). The coupling $\hat d
\cdot \vec E_0\cos(\Omega_L\tau)$  generates a factor
$\Omega_L$ that cancels the $1/\Omega_L$ in the renormalized cylindrical
Bessel function in Equation (7) of ref. \cite{NJP} in the Coulomb gauge, 
$\vec E(\tau)\,= - \frac{\partial}{\partial \tau} \vec A(\tau)$, that is written
in Fourier space as $\vec E(\Omega_L)\,=\,i\Omega_L\cdot\,\vec
A(\Omega_L)$. The Floquet sum, which is a consistency check, is discussed in
section \ref{sec:FLO}.

It has been shown by ref. \cite{Monien} that the
  coupling of an electromagnetic field modulation to the onsite electronic density 
  $n_i\,=\,c^{\dagger}_{i,\sigma}c_{i,\sigma}$ in the unlimited three
  dimensional translationally invariant system alone, can be gauged away. This
  type of coupling can be absorbed in an overall shift of the local potential
  while no additional dispersion is reflecting any additional functional dynamics of the system. Thus
  such a system \cite{Rigol,Bukov} will not show any topological effects as a topological
  insulator or a Chern insulator. In contrast, the coupling of the external
  electromagnetic field modulation to the dipole moment of the charges, and
  thus to the hopping term, see Equation (\ref{Hamilton_we}), as a kinetic energy
  of the fermions, cannot be gauged away and is causing the development of topological states in the three dimensional
  unlimited systems. A boundary as such is no necessary requirement. Line three of Equation (\ref{Hamilton_we}) formally represents
  the electromagnetically induced kinetic contribution
\begin{eqnarray}
\!\!\!\!\!\!\!\!\!\!\!\!\!\!
i\vec{d}\cdot\vec{E}_0 \cos(\Omega_L \tau)\sum_{<ij>,\sigma}\left(
           c^{\dagger}_{i,\sigma}c^{{\color{white}\dagger}}_{j,\sigma} 
 	  -
           c^{\dagger}_{j,\sigma}c^{{\color{white}\dagger}}_{i,\sigma}
         \right)\,=\,e\sum_{\vec{r}}\hat{j}_{ind}(\vec{r})\cdot\vec{A}(\vec{r},\tau)\label{curcont}
\end{eqnarray}

which is the kinetic contribution of the photo-induced charge
current in-space dependent with $\vec r$
\begin{eqnarray}
\vec{j}_{ind}(\vec{r})_{\delta}\,=\,-\,\frac{t}{i}\sum_{\sigma}(c^{\dagger}_{\vec
  r, \sigma}c_{\vec r +\delta, \sigma}\,-\,c^{\dagger}_{\vec r +\delta,
\sigma}c_{\vec r, \sigma}).
\end{eqnarray}
The temporal modulation of the
         classical external electrical
         field in (111) direction always causes a temporally
         modulated magnetic field contribution $\vec B (\vec
         r,\tau)\,=\,\nabla\times\vec{A}(\vec r,\tau)$ with $\vec B
         (\vec{r},\Omega_L)$ in Fourier space, as a
         consequence of Maxwells equations. In the following we derive
         the non-equilibrium local density of states (LDOS) which comes along
         with the dynamical life-time of non-equilibrium states as an inverse of the imaginary part of the
         self-energy $\tau\,\sim\,1/\Im\Sigma^R$. A time reversal
         procedure induced by an external field will never be able
         to revise the non-equilibrium effect. The photon-electron
         coupling and thus the absorption will be modified and the overall profoundly differing
         material characteristics is created. Conductivity and polarization
         of excited matter in the non-equilibrium are preventing any
         time-reversal process in the sense of closing the Floquet fan again
         in this regime. The initial electromagnetic field thus causes
         a break of the time-reversal symmetry, the current leads to the acquisition of a
         non-zero Berry flux. A Wannier-Stark type ladder
         \cite{Zak89} is created that can be characterized by it's Berry phase
         \cite{BerryElectronicProperties} as a Chern or a winding
         number or the $\mathbb{Z}_2$ invariants in three dimensions
         respectively \cite{Vanderbilt2017}.

\section{Floquet Spectra of Driven Semiconductors}

\label{sec:QUASIENERGY}

\begin{figure}
\vspace*{-2.0cm}
\hspace*{-0.8cm}\centering{\scalebox{1.2}{\includegraphics{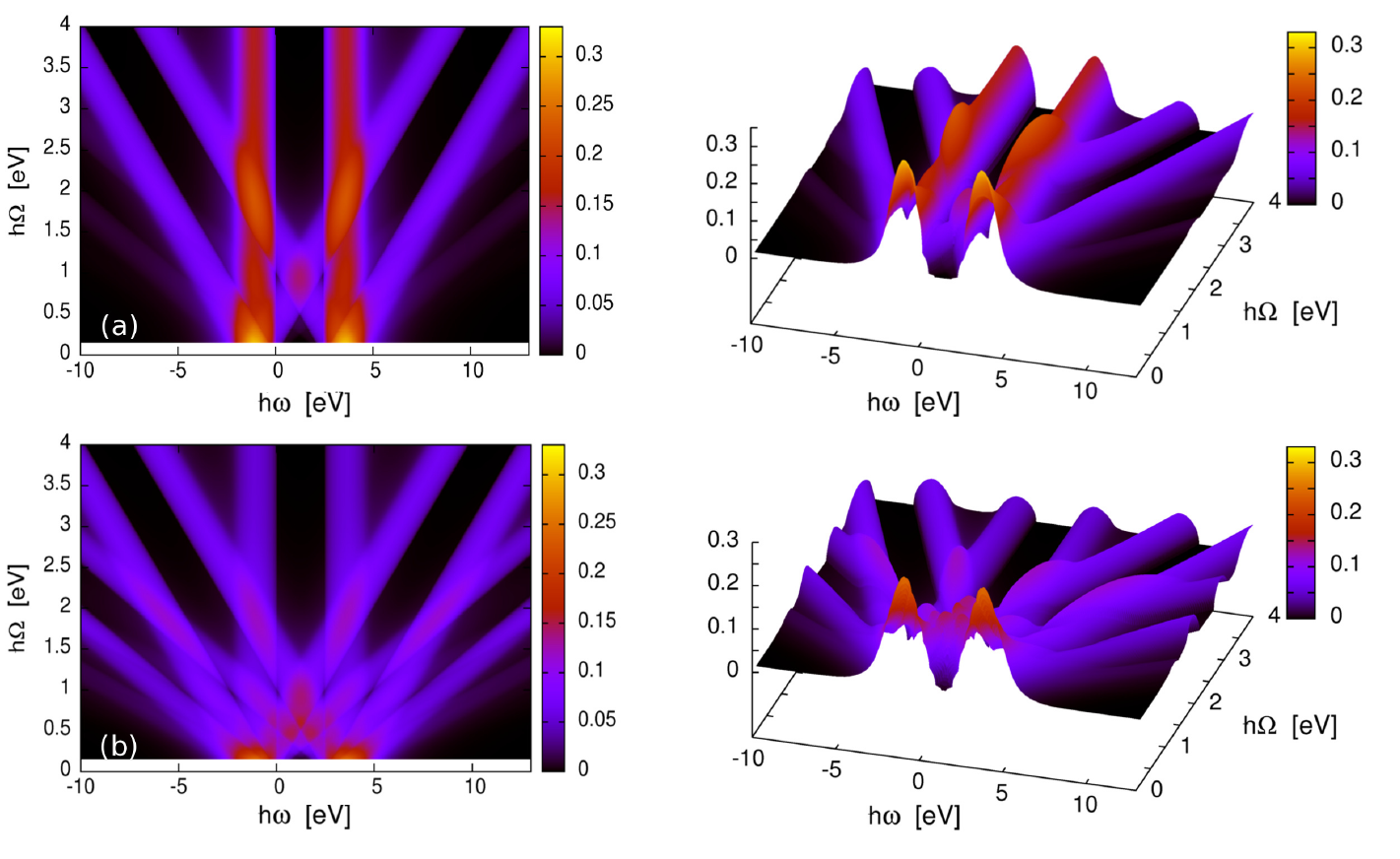}}}
\vspace*{0.5cm}     
\caption{Floquet topological quantum states of the
  semiconductor bulk in the non-equilibrium. (a) The evolution of the LDOS in the non-equilibrium for varying
  excitation laser frequencies $\Omega_L$ up to $\Omega_L\,=\,4.0\,eV$ is shown. The excitation
  intensity 5.0 $MW/cm^{2}$ is constant. The bandgap of ZnO 
  rocksalt in equilibrium is $2.45\,eV$, see Fig. \ref{Mott}, the gap is vanishing
  with the increase of the driving frequency
  and dressed states emerge as a consequence of  the non-equilibrium AC-Stark effect \cite{SchmittRink,Chemla}. The
  split bands are superposed by a doublet of Floquet fans which
  intersect. The formation of topological subgaps, see e.g. at
  $\hbar\Omega_L\,=\,0.9\,eV$ occurs. (b) The evolution of the LDOS for the excitation
  intensity of 10.0 $MW/cm^{2}$ is shown. Spectral weight is shifted to a
  multitude of higher order Floquet-bands, while the original split
  band characteristics almost vanishes apart from the near-gap band edges. A variety of Floquet gaps is
formed. At any crossing point topologically induced transitions are possible, the
generation of higher harmonics can be enhanced. Panels on the
right display the  topology of the LDOS. The subgaps are very pronounced and the
intersection of bands is visible as an increase of the LDOS which can be
measurable in a pump-probe experiment. For a detailed discussion please see section \ref{sec:QUASIENERGY}.}
\label{IMEffectiveGFD}       % Give a unique label
\end{figure}
\begin{figure}
\centering{\scalebox{1.5}{\includegraphics{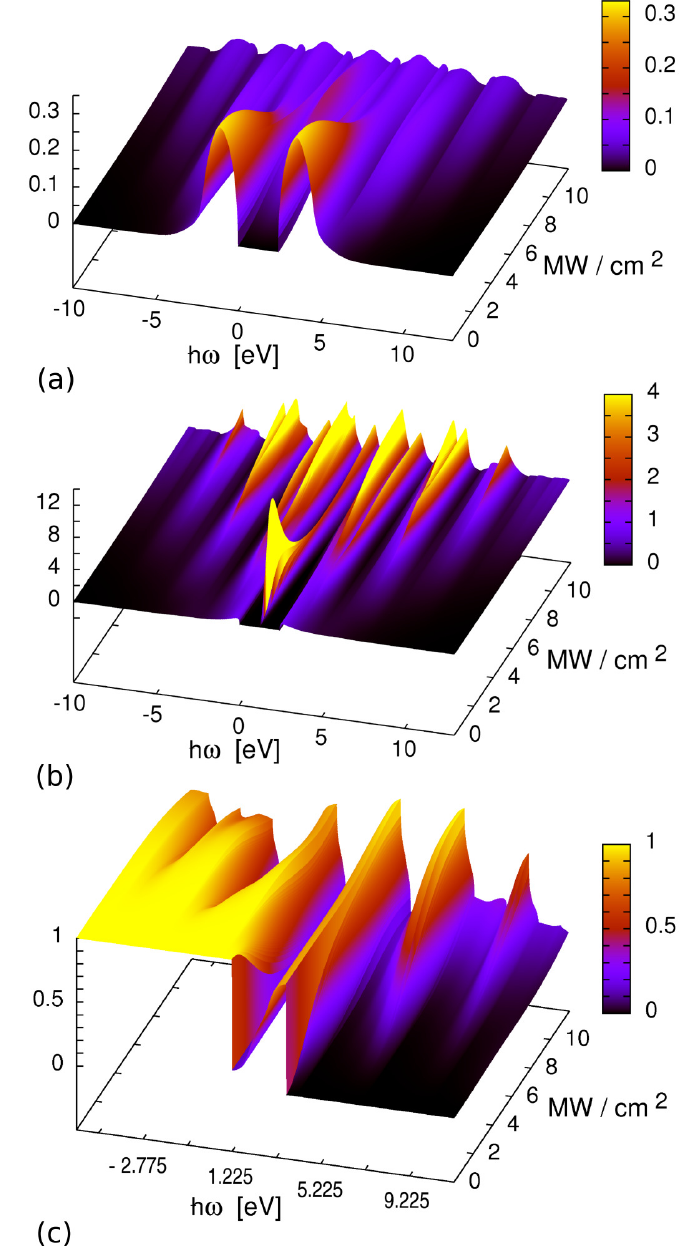}}}
%\vspace{0.5cm}       % Give the correct figure height in cm
\caption{(a) Energy spectra of Floquet topological quantum states of the
  semiconductor bulk in the non-equilibrium. The evolution of the LDOS is
  displayed for the single excitation energy of $\hbar\Omega_L=1.75\,eV$, wavelength
  $\lambda=710.0\,nm$ and an increasing external driving intensity up to $10.0\,MW/cm^2$. Spectral weight is shifted by excitation to Floquet
  sidebands and a sophisticated sub gab structure is formed. In the non-equilibrium such
  topological effects in correlated systems are non-trivial. The
  bandgap in equilibrium is $2.45\,eV$, the Fermi edge is $1.225\,eV$, the width
  of each band is $2.45\,eV$ as well. (b) Inverse lifetime $\Im\Sigma^R$ of Floquet states
  of electromagnetically driven ZnO  rocksalt bulk in the
  non-equilibrium. (c) Non-equilibrium distribution function $F^{neq}$ of electrons in optically driven
  bulk ZnO rocksalt. Parameters in (b) and (c) are identical to (a). For a detailed discussion please see
  section \ref{sec:QUASIENERGY}. }
\label{IMEffectiveGAD}       % Give a unique label
\end{figure}

From the numerically computed components of the Green's function, 
we define \cite{NJP} the local density of states (LDOS), $N(\omega,\Omega_L)$, where  momentum is integrated out and Floquet indices are summed
\begin{eqnarray}
N(\omega,\Omega_L)
=
-\frac{1}{\pi}\sum_{mn} \int {\rm d^3}k  {\rm Im\,} 
G^{R}_{mn} ({\bf k},\omega, \Omega_L).
\label{Eq:DoS}
\end{eqnarray}
In combination with the lifetime as the inverse of the imaginary part of the
self-energy, $\tau\,\sim\,1/\Im\Sigma^R$, and the non-equilibrium distribution
function 
\begin{eqnarray}
F^{neq}(\omega,\Omega_L)=
\frac {1}{2}
\left(
1
+
\frac{1}{2i}
\frac{\sum_m G_{0m}^{Keld}(\omega,\Omega_L)}{\sum_n{ \rm Im}G_{0n}^{A}(\omega,\Omega_L)}
\right).
\label{Eq:FNEQ}
\end{eqnarray}
the local density of states
$N(\omega,\Omega_L)$ can be experimentally determined as the compelling
band structure of the non-equilibrium system. 

We show results for optically excited semiconductor bulk, with a band gap in the equilibrium of $2.45\,eV$ and typical parameters
for ZnO. ZnO in either configuration \cite{JiaLu1,JiaLu2,ZnOMono1,ZnOMono2,Blaaderen} is a very promising material for the construction of
micro-lasers, quantum wells and optical components. In certain geometries and in
connection to other topological insulators it is already used for the
engineering of ultrafast switches. ZnO, see Fig. (\ref{ZSTRUCT}), is broadly
investigated in the  non-centro-symmetric wurtzite configuration and very
recently in
the centro-symmetric rocksalt configuration
\cite{Rocksalt1,Rocksalt2}, it's bandgap is estimated to be of 1.8 eV up to 6.1 eV depending on various factors
as the pressure during the fabrication process. In either crystal configuration the production of
second or higher
order harmonics under intense external excitations \cite{Wegener} is
searched. It is of high interest for novel types of lasers.

\subsection{Development and Lifetimes of Floquet Topological Quantum States in the
  Non-Equilibrium}
\label{sec:FLO}

In Fig. \ref{IMEffectiveGFD} we investigate a wide gap semi-conductor band structure,
the band gap in equilibrium is assumed to be $2.45\,eV$. The semiconductor
bulk  shall be exposed to an
external periodic-in-time driving field. The system is so far
considered as pure bulk, so we are investigating  Floquet topological
effects in the non-equilibrium without any other geometrical influence. The
excitation intensity in the results of Fig. \ref{IMEffectiveGFD}(a)  is
considered to be $5.0 MW/cm^2$  and $10.0
MW/cm^2$ in Fig. \ref{IMEffectiveGFD}(b). DMFT as
  a solver for correlated and strongly correlated electronics as such is a
spatially independent method. It is designed to derive bulk effects,
whereas all k-dependencies have been
integrated as the fundamental methodology. Therefore we are not analyzing the k-resolved information of the
Brillouin zone. As long as no artificial coarse graining with a novel
length scale in the sense of finite elements or finite volumes is included, DMFT results in one, two and three dimensions are independent of
any spatial information. In fact however the energy dependent LDOS profoundly
changes with a varying excitation frequency and with a varying excitation intensity
as well, which gives evidence that also the underlying k-dependent band structure is
topologically modulated. A non-trivial topological structure of the Hilbert
space is generated by external excitations even though our system in equilibrium is fully
periodic in space and time. The time dependent external electrical field
generates a temporally modulated magnetic field which results in a dynamical
Wannier-Stark effect and the generation of Floquet states. Floquet states are the temporal
analogue to Bloch states, and thus the argumentation by Zak
\cite{Zak89} in principle applies for the generation of the Berry phase $\gamma_m$, since the
solid is exposed to an externally modulated electromagnetic potential
\cite{BerryElectronicProperties,Vanderbilt1,Vanderbilt2,Resta}. The Floquet quasi-energies, see
Fig. \ref{FloquetGreen}, are labeled by the Floquet modes in dependency to the external excitation frequency,
and to the external excitation amplitude. The topological invariants, the
Chern number as a sum over all
occupied bands $n\,=\,\sum_{m=1}^{\nu}n_m \neq 0$ and the $\mathbb{Z}_2$
invariants include the Berry flux $n_m\,=\,1/2\pi\int d^2 \vec{k} (\nabla
\times \gamma_m)$. The winding number is also consistently  associated with the argument of
collecting a non-zero Berry flux. We consider both regimes, where the driving frequency is smaller than the width of the semiconductor gap in
equilibrium and also where it is larger and a very pronounced topology of
states is generated.  While the system is excited and thus evolving in
  non-equilibrium, a Berry phase is acquired and a non-zero Berry flux and
  thus a non-zero Chern number are characterizing the topological
  band structure as to be {\it non-trivial}. 
For one dimensional models \cite{Zak89,DalLago} with the variation of the
external excitation frequency $\Omega_L$ replica of Floquet bands with a
quantized change of the Berry phase $\gamma\,=\,\pi$ emerge in the spectrum. In three
  dimensions 
  the Berry phase is associated with the Wyckoff positions of the crystal and the Brillouin
  zone \cite{Zak89}, and as such it cannot be derived by
  the pure form of the DMFT. $\vec k$-dependent information can be derived
  by so called real-space or cluster DMFT solutions (R-DMFT or CDMFT)
  \cite{Gull,Hofstetter,Peters1,Peters2}, however they have not been
  generalized  to the non-equilibrium for three-dimensional systems. Important
  to note is that the system out of
  equilibrium acquires a non-zero Berry phase and Coulomb interactions lead to
  a Mott-type gap which closes due to the superposition by crossing Floquet bands, however also the opening of non-equilibrium induced Mott-gap
  replica can be found for $\Omega_L\,=\,0.95\,eV$. The replica are complete at $\Omega_L\,=\,1.9\,eV$, see
Fig. \ref{IMEffectiveGFD}(a). For the increase of $\vec E_0$ these gap replica are
again intersected by next order of Floquet sidebands. The
closing of the Mott-gap and the opening of side Mott-gaps, in the spectrum due to topological excitation are
classified as {\it non-trivial} topological effects.  In
Fig. (\ref{IMEffectiveGAD}(a)) we present results of the LDOS for the same
system of optically excited cubic ZnO rocksalt excited by an external laser
energy of $1.75\,eV$ and an increasing excitation intensity,
Fig. (\ref{IMEffectiveGAD}(b)) shows the corresponding inverse lifetime $\Im
\Sigma^R$ and Fig. (\ref{IMEffectiveGAD}(c)) shows the corresponding
non-equilibrium distribution of electrons $F^{neq}$. Also for very  small
excitation intensities the result for the non-equilibrium distribution
function $F^{neq}$ shows a profound deviation from the Fermi step in
equilibrium. These occupied non-equilibrium states have a finite life time
especially at the inner band-edges which is a sign of the Franz-Keldysh effect
\cite{Franz,Keldysh}, here in the sense of a topological effect, which is
accessible in a pump-probe experiment.\\ The change of the
  polarization of the external excitation modifies the physical situation and the result. Especially
  circular and elliptically polarized light can be formally written as a
  superposition of linear polarized waves. So, in the pure uncorrelated case,
  $U\,=\,0$, one could think that the setup can be formally implemented in the sense of coupled matrices.
In the strongly correlated system at hand the physics is fundamentally
different. The solution for the
strongly correlated case, $U\,\neq\,0$, in the non-equilibrium, including
DMFT, will become more sophisticated since the coupled matrices will result in
the entanglement of processes in some sense. This can be deduced from the
result in Fig. (\ref{IMEffectiveGAD}(b)) which displays the modification of
non-equilibrium life-times of electronic states due to the varying excitation
amplitudes. Such a modification is also qualitatively found for varying
excitation frequencies.\\
The classification of correlated topological systems is an active research
field \cite{Rigol,Manmana,Rachel}. At this point we refer to the section
  \ref{sec:FLO} in this article, where we show that in our theoretical
  results the analysis of the single Floquet modes.
For the investigation of the LDOS and the occupation
number $F^{neq}$, as well as the life-times of the non-equilibrium states, an artificial cut-off of the Floquet
series, as it is described in the literature, does not make sense from the
numerical physics point of view of DMFT in frequency space. This would hurt
basically conservation laws and the
cut-off would lead to a drift of the overall energy of
the system, see section \ref{sec:FLO}. However according to the bulk-boundary
correspondence \cite{KaneMele,HasanKane} the results of this work for bulk
will be observed in a pump-probe experiment at the surface of the
semiconductor sample. In the following we discuss the development of
  Floquet topological states in the correlated system for an increasing external driving frequency
  $\Omega_L$, see Fig.(\ref{IMEffectiveGFD}), and for an increasing amplitude of
  the driving, see Fig.(\ref{IMEffectiveGAD}).

\subsection{Topological Generation of Higher Harmonics and of Optical Transparency}
\label{sec:HHG}

When we increase
the external excitation energy of the system $\Omega_L$ from $0\,eV$ to
$4.0\,eV$, Floquet topological quantum states as well as the topologically
induced Floquet band gaps for bulk matter are developed. Both, valence and
  conduction band split in a multitude of Floquet sub-bands which cross each
  other.  In Fig. \ref{IMEffectiveGFD}(a) the evolution of
a very clear Floquet fan for the valence as well as for the conduction band of the
correlated matter in the non-equilibrium is found.  When the excitation energy
is increased up to $0.45\,eV$ the original band gap is
  subsequently closing, the first crossing point ins the semiconductor gap
  along the Fermi edge is found at $0.45\,eV$. With the increase of the excitation
  intensity, see Fig. \ref{IMEffectiveGFD}(b), higher order Floquet sidebands
  are gaining spectral weight and we find the next prominent crossing point at the Fermi-edge for
  $0.2\,eV$. Band edges of higher order Floquet bands form crossing points
  with those of the first order. For an excitation energy of  $0.42\,eV$
  the crossing points of the first side bands (02) with the higher number side
  bands are found at the atomic energy of $1.08\,eV$ and $2.3\,eV$, so above the
  valence band edge and deep in the gap of the
  semiconductor. Semiconductors are well known for fundamental absorption at
  the band edge of the valence band. We find here that the absorption
  coefficient of the semiconductor is topologically modulated. Non-trivial
  transitions at the crossing points of Floquet-valence subbands and
  Floquet-conduction subbands become  significant. A higher order Floquet subband is usually physically reached by absorption or
  generation  of higher harmonic procedures and  we find a high
  probability for a topologically induced direct transitions from the
  fundamental to higher order bands for those points in the spectrum where a
  Floquet band-edge intersects with the
  inner band edge of the equilibrium valence band. At any bandedge directional
  scattering can be expected if the life-times of states are of a value that
  is applicable to the expected scattering processes.
In general the optical refractive index is topologically modulated, electromagnetically induced
transparency will become observable for intense excitations. The
topologically induced Floquet bands overlap and cross each other. Consequentially
very pronounced features and narrow subgaps are formed
in the LDOS which correspond with sharp spikes in the expected life-times in
the non-equilibrium. Floquet replica of valence and
conduction bands are formed and the dispersion is renormalized. We also find
regions for excitation
energies from $\hbar\Omega_L=0.5\,eV$ up to $\hbar\Omega_L=0.85\,eV$ and from
$\hbar\Omega_L=1.1\,eV$ up to $\hbar\Omega_L=1.45\,eV$ which can be
interpreted as  a topologically
induced metallic phase. These states are the result of the Franz-Keldysh
effect \cite{Franz,Keldysh} or AC-Stark effect which is well known for high intensity excitation
of semiconductor bulk and quantum wells \cite{SchmittRink,Chemla}. \\ 
From the viewpoint of correlated electronics in the
non-equilibrium we interpret our results as follows. For finite
excitation frequencies an instantaneous transition to the topologically
induced Floquet band structure and a renormalized dispersion is derived. In the bulk system clear
Floquet bands develop, if the sample is excited by an intense
electrical field. This is observable in  Fig. \ref{IMEffectiveGFD}. \\
In Fig. \ref{IMEffectiveGAD} we display the same system as in  Fig. \ref{IMEffectiveGFD}
for constant driving energy of $1.75\,eV$ and an increasing driving
intensity up to $10.0\, MW/cm^2$. We find the development of side bands and an overall
vanishing semiconductor gap is found, which marks the transition from the semi-conductor to the topologically
highly variable and switchable conductor in the non-equilibrium.\\
In this article we do not investigate the coupling to a geometrical edge or a resonator mode. This will lead in the optical case to additional contributions in
Equation (\ref{Hamilton_we}) for the mode itself $\hbar\omega_o
a^{\dagger}a^{{\color{white}\dagger}}$  and the coupling term of the resonator or
edge mode  to the electron system of the bulk $g\!\sum_{i, \sigma}
c^{\dagger}_{i,\sigma}c^{{\color{white}\dagger}}_{i,\sigma}(a^{\dagger}\! +
a)$. $a^{\dagger}$ and $a$ are the creator and the annihilator of the
photon, $g$ is the variable coupling strength of the photonic mode to the
electronic system \cite{Lamata}. From our results here we can conclude already that for
semi-conductor cavities and quantum wells as well as for structures which
enhance so called edge states, these geometrical edge or surface resonances will induce an
additional topological effect within the full so far excitonic spectrum. It is
an additional effect that occurs beyond the bulk boundary correspondence. Dressed states may release energy quanta,
e.g. light, or an electronic current into the resonator component \cite{PRB}. So we
expect from our results that such modes may become a sensible switch in non-equilibrium. \\
It can be expected as well that novel topological effects in the
non-equilibrium occur from the geometry. If the energy of the system is
conserved, these modes will have always an influence on the full spectrum
of the LDOS, when the system is otherwise
periodic in space and time. Thus it is to clarify  whether such modes may be of
technological use.  For the investigation of ZnO as a laser material the influences of surface resonators will be subject to
further investigations. It is on target to find out all the signatures of
a topologically protected edge mode  in correlated and strongly correlated
systems out of equilibrium, and  to classify the significance of topological
effects  for the occurrence of the electro-optical Kerr effect, the magneto-optical Kerr effect (MOKE) or the surface magneto-optical
Kerr effect (SMOKE). We believe that in correlated many-body
  systems out of equilibrium a bulk boundary correspondence is given and will
  be experimentally found. Those results  become
modified or enhanced by a coupling of bulk states with the geometry of a
micro- or a nanostructure and their geometrical resonances.

\subsection{Consistency of the Numerical Framework}
\label{sec:FLO}

The consistency of the numerical formalism is generally checked by the sum over
all Floquet indices with the physical meaning that energy conservation must be
guaranteed in the non-equilibrium. Consequentially we do not take into account
thermalization procedures and the systems temperature remains constant. The
analysis of the numerical validity as the normalized and frequency integrated density of states 
\begin{eqnarray}
N_{i}(\Omega_L)
:=
\int{\rm d}\omega N(\omega,\Omega_L)=1
\label{Eq:N_integrated}
\end{eqnarray}
\begin{figure}
\vspace*{-1.1cm}\centering\resizebox{0.56\textwidth}{!}{%
  \includegraphics{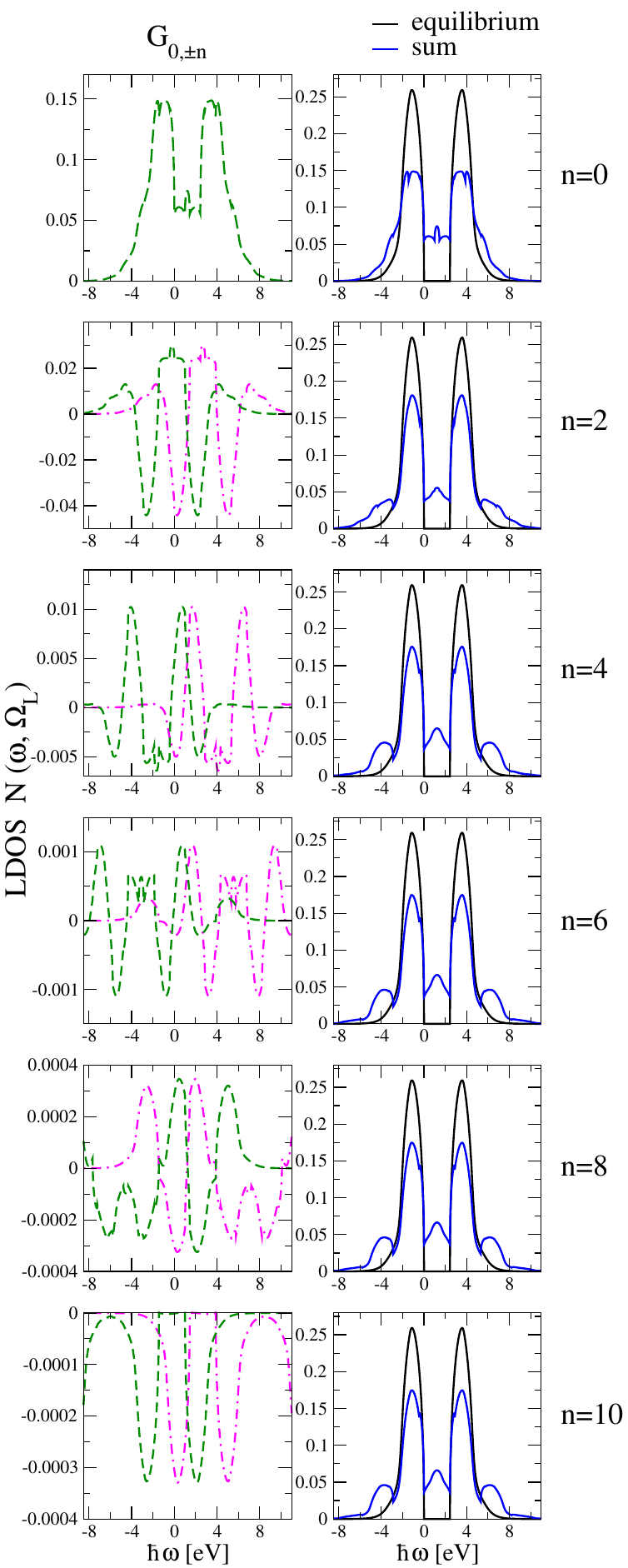}}
\vspace{0.5cm}       % Give the correct figure height in cm
\caption{Floquet contributions and accuracy check of the numerical results for
  the LDOS of driven semiconductor bulk. The bandgap in equilibrium is $2.45\,eV$, the Fermi edge is $1.225\,eV$. For
  discussion please see section \ref{sec:FLO}.}
\label{Floquet_Panel}       % Give a unique label
\end{figure}
is confirmed in this work for summing over Floquet indices up to the order of
10. We discuss in Fig. \ref{Floquet_Panel} on the l.h.s. the Floquet
contributions with increasing number in steps of
$n=0,2,4,6,8,10$. With an increasing order of the Floquet index the amplitude of
the Floquet contribution decreases towards the level of numerical precision of
the DMFT self-consistency. This is definitely reached for $n=10$
and thus it is the physical argument to cut the Floquet expansion off for
$n=10$. As a systems requirement the Floquet contributions $G_{0\pm n}$ are perfectly mirror symmetric with respect to the Fermi edge, whereas the sum of
both contributions is directly symmetric with respect to the Fermi edge. These
symmetries are generally a proof of the validity of the numerical Fourier
transformation and the numerical scheme. The order of magnitude of each Floquet
contribution with a  higher order than  $n=4$ is almost falling consistently with the
rising Floquet index. We display results for the external laser wavelength of
$\lambda=710.0\,nm$ and the laser intensity of $3.8\,MW/cm^2$, the ZnO gap is
assumed to be $ 2.45\,eV$ which is ZnO rocksalt as a laser active material. We include
Floquet contributions up to a precision of $10^{-3}$ with regard to their
effective difference from the final result on the r.h.s of
Fig. \ref{Floquet_Panel} as the sum to the $n$th-order. It corresponds to
the accuracy of the self-consistent numerics. \\
The Floquet contributions, Fig. \ref{Floquet_Panel},  as such consequentially do
not have a direct physical interpretation however the sum of all contributions
is the local density of states, the LDOS, as a material
characteristics. Whereas the lowest order Floquet contribution, compare Equation (\ref{Floquet-Fourier}),
$G_{00}$ is symmetric to the Fermi edge but strictly positive, higher order
contributions $G_{0\pm n}$ are mirror symmetric to each other and in sum they
can have negative contributions to the result of the LDOS. The order of the Floquet contribution $n$ numbers the evolving
Floquet side bands which emerge in the LDOS, compare Fig. \ref{IMEffectiveGFD}(a). The increase of mathematical and
numerical precision  has direct consequences for the finding and the accuracy of physical
results, and the investigation of the coupling of the driven electronic
system of the bulk with edge and surface modes will therefore profit. Bulk-surface coupling effects in nanostructure and waveguides are of
great technological importance and the advantage of this numerical approach in
contrast to time dependent DMFT frameworks in this respect is 
obvious.

\section{Conclusions}

We investigated in this article the development of Floquet topological quantum
states in wide band gap semiconductor bulk as a correlated
electronic system with a generalized Hubbard model and with dynamical mean field
theory in the non-equilibrium. We found that optical excitations
induce a non-trivial band structure and  in several frequency ranges a
topologically induced metal phase is found as a
result of the AC-Stark effect. The intersection of Floquet bands and
band-edges induces novel transitions,  which may lead to up- and
downconversion effects as well as to higher harmonic
generation. The  semiconductor absorption
  coefficient is topologically modulated. Non-trivial
  transitions at the crossing points of the underlying equilibrium
  band structure with the intersecting Floquet fans become
  possible and their efficiency is depending on the excitation power. We also find the development of pronounced novel sub gaps as areas of
electromagnetically induced transparency.  We also presented a consistency
check as a physical consequence of the Floquet sum which ensures numerically
energy conservation. Our results for semiconductor bulk  can be tested 
optoelectronic and magneto-optoelectronic experiments, they may serve as
a guide towards innovative
laser systems. The bulk semiconductor under topological non-equilibrium excitations as such has to be reclassified. It will be of great interest to investigate the interplay of topological bulk effects with additional surface resonances, a polariton coupling, or a surface magneto-optical modulation.

% \bibliographystyle{}
% \bibliography{}

%%%%%%%%%%%%%%%%%%%%%%%%%%%%%%%%%%%%%%%%%%
\vspace{6pt} 

%%%%%%%%%%%%%%%%%%%%%%%%%%%%%%%%%%%%%%%%%%
%% optional
%\supplementary{The following are available online at \linksupplementary{s1}, Figure S1: title, Table S1: title, Video S1: title.}


%%%%%%%%%%%%%%%%%%%%%%%%%%%%%%%%%%%%%%%%%%
\authorcontributions{Both authors equally contributed to the presented
  work. Both authors were equally involved in the preparation of the
  manuscript. Both authors have read and approved the final manuscript.}

\funding{This research received no external funding.}

%%%%%%%%%%%%%%%%%%%%%%%%%%%%%%%%%%%%%%%%%%
\acknowledgments{The authors thank H.Monien, H.Wittel and F.Hasselbach for
  highly valuable discussions.}

\conflictsofinterest{No competing interests.}


%=====================================
% References, variant A: internal bibliography
%=====================================
\reftitle{References}

% The following MDPI journals use author-date citation: Arts, Econometrics, Economies, Genealogy, Humanities, IJFS, JRFM, Laws, Religions, Risks, Social Sciences. For those journals, please follow the formatting guidelines on http://www.mdpi.com/authors/references
% To cite two works by the same author: \citeauthor{ref-journal-1a} (\citeyear{ref-journal-1a}, \citeyear{ref-journal-1b}). This produces: Whittaker (1967, 1975)
% To cite two works by the same author with specific pages: \citeauthor{ref-journal-3a} (\citeyear{ref-journal-3a}, p. 328; \citeyear{ref-journal-3b}, p.475). This produces: Wong (1999, p. 328; 2000, p. 475)

%=====================================
% References, variant B: external bibliography
%=====================================
%\externalbibliography{yes}
%\bibliography{your_external_BibTeX_file}

%%%%%%%%%%%%%%%%%%%%%%%%%%%%%%%%%%%%%%%%%%
%% optional
%\sampleavailability{Samples of the compounds ...... are available from the authors.}

%% for journal Sci
%\reviewreports{\\
%Reviewer 1 comments and authors’ response\\
%Reviewer 2 comments and authors’ response\\
%Reviewer 3 comments and authors’ response
%}

%%%%%%%%%%%%%%%%%%%%%%%%%%%%%%%%%%%%%%%%%%
\end{document}